\documentclass[a4paper,fleqn,usenatbib]{mnras}

\usepackage{txfonts}

\usepackage[T1]{fontenc}
\usepackage{ae,aecompl}

\usepackage{graphicx}
\usepackage{multirow}
\usepackage{comment}
\usepackage{amssymb}
\usepackage{wasysym}
\usepackage{color}
\usepackage{hyperref}

\title[The post-outburst pulsations of GW Lib]{The post-outburst pulsations of the 
       accreting white dwarf in the cataclysmic variable GW Librae}

\author[P.~Chote, D.~J.~Sullivan]
       {P.~Chote$^{1,2}$\thanks{E-mail: p.chote@warwick.ac.uk}\footnotemark[3], 
        D.~J.~Sullivan$^{2}$\thanks{Email: denis.sullivan@vuw.ac.nz}\thanks{Visiting Astronomer, University of Canterbury Mt John Observatory}\\
$^1$Department of Physics, University of Warwick, Coventry CV4 7AL, United Kingdom\\
$^2$School of Chemical \& Physical Sciences, Victoria University of Wellington, 
    P.O. Box 600, Wellington 6104, New Zealand}

\begin{document}

\date{Accepted 2016 February 19}

\pubyear{2016}

\pagerange{\pageref{firstpage}--\pageref{lastpage}}
\maketitle

\label{firstpage}

\begin{abstract}
 We present new time-series photometry of the accreting pulsating white
 dwarf system GW Librae obtained in 2012 and 2013 at the University
 of Canterbury Mt John Observatory in New Zealand. Our 2012 data show the return of a $\sim$19
 minute periodicity that was previously detected in 2008.
 This pulsation mode was a dominant feature of our quality May 2012
 data set, which consisted of six contiguous nights; a detailed
 analysis indicated a degree of frequency variability.
 We show by comparison with the previously identified pulsation modes that
 this periodicity is best explained as a new mode, and that the
 quasi-stability of the periods appears to be a general feature of the
 pulsations in these systems. We also find a previously unreported 3-hour
 modulation period, which we believe to be related to the known two and four
 hour periods of so far unknown origin.
\end{abstract}


\section{Introduction}
White dwarfs (WDs) are the final evolutionary state of around 98 per cent of
stars in our Galaxy. Efficient chemical diffusion in their envelopes
due to the high surface gravities produces atmospheres composed
predominantly of the lightest elements hydrogen and helium. About 80 per cent
of WDs have a hydrogen atmosphere (spectral type DA) overlaying a
helium zone, and the bulk of the remaining 20 per cent (spectral type DB)
have helium atmospheres.

As isolated WDs slowly cool they pass through temperature regimes in which
nonradial $g$-mode pulsations are observable. For the DA WDs the
instability strip occurs near T$_{\rm eff} \sim 12$\,kK in which a
partially ionized H layer in the envelope drives the pulsations
\citep{dziembowski81, brickhill91}. The DB instability strip occurs
in a significantly higher and less well-defined temperature regime
(T$_{\rm eff} \sim 23 - 28$\,kK) as the driving mechanism involves a
partially ionized He\,\textsc{i} layer in the outer regions
\citep{winget82}. This temperature uncertainty is primarily due to the
difficulty of establishing a T$_{\rm eff}$ value based on the largely
temperature-independent behaviour of the He\,\textsc{i} spectral lines in the
temperature interval \citep{bergeron11, chote13}.

The nonradial pulsations are of low spherical harmonic order, have
periods $\sim 10^{2} - 10^{3}$\,s, and involve largely horizontal
material displacements along gravitational equipotential surfaces.
This creates surface temperature oscillations, leading to observable
flux variations.

The eigenfrequencies of these pulsations provide a fingerprint of the
internal structure of the WD, and if enough modes are detected
asteroseismic techniques can be employed to extract measurements 
of basic properties such as the size of the degenerate core of the WD,
rotation rates, and the mass, depth, and convection zone properties
of the helium and hydrogen layers in their envelopes \citep[see e.g.]
[for comprehensive reviews on the field]{winget08, fontaine08, althaus10}.

Both the period and shape of the luminosity variations associated with
a pulsation mode change as a function of temperature
\citep{mukadam06}, but because isolated WDs evolve very slowly through
the instability strips, we are limited to probing a single snapshot in
an individual star's evolution. The DAVs evolve at the slowest rate
\citep[\mbox{dP/dt $\sim10^{-15}\,$s~s$^{-1}$:} e.g.][]{kepler05,
mukadam13, sullivan15}, while the hotter DBVs, unsurprisingly, are
expected to evolve more rapidly \citep{winget04}.

A small number of systems, of which GW Librae (GW~Lib) is the prototypical
member, feature a pulsating WD that is accreting material from a close companion
star via an accretion disk. These systems (which exist as a sub-group
of the Cataclysmic Variables) undergo outbursts when the
accretion disk reaches a critical density, becomes thermally unstable,
and rapidly dumps material onto the WD. The conversion of gravitational
potential energy to heat raises the surface temperature of the WD to
a few thousand Kelvin above the instability strip, and the pulsations cease.
Over the following few years the pulsations return and evolve as the
WD cools back towards its pre-outburst temperature \citep{piro05}.

These systems have a low accretion rate, and so the WD dominates the visible
and ultraviolet luminosity (estimated as $\gtrsim$ 60 per cent for GW Lib in the $V$
band \citep{vican11}) when the system is not undergoing an outburst. This
makes it possible to detect the intrinsic luminosity variations generated by
the pulsations during the long quiescent periods between outbursts
(\mbox{$\gtrsim 10\,$years}).

A side effect of the accretion is that the WD envelope composition is
contaminated with helium and other elements from the companion star.
This creates a more complicated parameter space in which pulsations
may be driven over a wider range of temperatures than the traditional
DA and DB instability strips \citep{arras06, vangrootel15}.
The accumulated angular momentum spins the WD up to fast rotation rates
(down to a few minutes), which will have an as yet undetermined
effect on the pulsations, and is relevant for the discussion of the
fast rotating `super-chandrasekhar' WDs that have been proposed as Type Ia
supernovae candidates \citep[e.g.][]{yoon05}.

GW Lib was the first accreting pulsating WD to be identified, and
remains to date the most well studied. Subsequent discoveries have
provided $>15$ additional systems (see Table 6 of
\cite{szkody10}, plus the more recent discoveries by \cite{pavlenko09,
woudt11, uthas12}). The historical time line of the key observational
programmes targeting GW Lib and their main results are summarised in
Table \ref{table:history}.

\begin{table*}
\label{table:history}
\centering
\caption{A brief tabular history of the key observational papers studying GW Librae.}
\begin{tabular}{cclr}
\hline
 \multicolumn{2}{c}{Observation Date} & \multicolumn{1}{c}{Observations} & Reference \\
\cline{1-2} Year & Month \\
\hline
1983 & August  & First outburst detected: mag $\sim18 \longrightarrow\ \sim9$ (nova classification) & \cite{maza83} \\
1987 & --      & Spectroscopic classification as dwarf Nova (DN)                                    & \cite{duerbeck87} \\
1991 & June    & Spectroscopic classification as DN of WZ Sge type                                  & \cite{ringwald96} \\
1997 & March   & Pulsations in white dwarf primary discovered near 390, 650\,s                      & \cite{warner98} \\
1998 & April   & Spectroscopic measurement of 76.4 min orbital period                               & \cite{szkody00} \\
1998 & May     & Multi-site photometry campaign: pulsation periods near 650, 370, 230\,s            & \cite{vanzyl04} \\
1999 & June    & Refined 76.79 min spectroscopic orbital period determination                       & \cite{thorstensen02} \\
2001 & May     & Discovery of an unknown 2.1\,h periodicity.                                        & \cite{woudt02} \\
2002 & January & HST observations show large UV pulsation amplitudes and solar-like metals          & \cite{szkody02} \\
2002 & May     & Spectroscopic measurement of 97\,s spin period.                                    & \cite{vanspaandonk10} \\
2005 & May     & VLT multicolour observations clearly show pulsations and 2.1\,h modulation         & \cite{copperwheat09} \\ 
\hline
2007 & April   & Second outburst detected: peaked at 8 mag                                          & \cite{templeton07} \\
$\cdots$ &     & Photometric observations by several groups following...                            & \cite{vican11} \\
$\cdots$ &     & ...the evolution of the outburst and subsequent cooling                            & \cite{kato08} \\
2007 & April   & Swift observations show unusual x-ray flux increase during outburst                & \cite{byckling09} \\
2007 -- 2009 & & Some photometric variability observed, including a new $\approx$19\,min period...  & \cite{copperwheat09} \\
$\cdots$ &     & ...plus 2.1\,h \& 4\,h modulations.  No clear evidence for WD pulsations           & \cite{vican11} \\
$\cdots$ &     &                                                                                    & \cite{bullock11}\\
2010 & March   & HST observations detect pulsations                                                 & \cite{szkody12}  \\
2011 & April   & Pulsations detected via HST and ground-based observations                          & \cite{szkody12}  \\
2012 & May     & Photometric study of the 19 minutes WD pulsation                                   & This paper \\
2013 & May     & Spectroscopic study of a 3.85\,h modulation                                        & \cite{toloza15} \\
\hline
\end{tabular}
\end{table*}

\begin{table}
\label{table:runs}
\centering
\caption{Table of GW Lib observation runs. All runs were acquired using a BG40
filter with the Puoko-nui instrument on the 1\,m telescope at UCMJO.}
\begin{tabular}{rrcrr}
\hline
\multicolumn{1}{c}{UT Date}   & 
\multicolumn{1}{c}{UT Start}  & 
\multicolumn{1}{c}{Exposure}  &
\multicolumn{1}{c}{Useable}   &
\multicolumn{1}{c}{0.1\% FAP} \\

& & \multicolumn{1}{c}{(sec)} &
\multicolumn{1}{c}{Hours} &
\multicolumn{1}{c}{(mma)}\\
\hline
 2 Mar 2011 & 13:09:00 & 20 & 4   & 10.4 \\
 4 Mar 2011 & 14:33:00 & 20 & 1.5 & 11.8 \\
\hline
 1 Jul 2011 & 07:02:00 & 20 & 5   & 4.9 \\
 2 Jul 2011 & 06:23:00 & 20 & 4.5 & 4.3 \\
 4 Jul 2011 & 06:36:00 & 20 & 7.5 & 3.8 \\
 6 Jul 2011 & 06:35:00 & 20 & 3   & 6.9 \\
\hline
27 Jul 2011 & 06:48:00 & 20 & 5.5 & 5.8 \\
 1 Aug 2011 & 06:39:00 & 20 & 5   & 5.3 \\
 2 Aug 2011 & 06:35:00 & 20 & 3   & 5.8 \\
\hline
24 Mar 2012 & 13:02:00 & 30 &  3.5 & 8.4 \\
25 Mar 2012 & 12:40:00 & 30 &  5.0 & 7.2 \\
23 Apr 2012 & 11:22:00 & 30 &  7.0 & 5.5 \\
\hline
17 May 2012 &  6:53:00 & 30 & 11.0 & 6.1 \\
18 May 2012 &  6:44:00 & 30 & 11.0 & 6.8 \\
19 May 2012 &  6:33:00 & 30 & 11.5 & 5.2 \\
20 May 2012 &  7:38:30 & 30 &  6.0 & 7.6 \\
21 May 2012 &  7:31:30 & 30 & 10.0 & 6.2 \\
22 May 2012 &  6:28:30 & 30 & 12.0 & 5.7 \\
\hline
13 March 2013 & 12:46:00 & 30 & 4.7 & 5.5 \\
\hline
\end{tabular}
\end{table}

\section{Mt John observations}

In 2011 we joined an observational programme to monitor GW Lib as it cooled
from its 2007 outburst. We obtained nine runs in 2011 which were originally
reported in \cite{szkody12}. The primary focus of this paper
are an additional nine runs that were obtained between March and May 2012
(initially reported in \cite{chote13b}) plus an additional run obtained in
March 2013. The details of these 19 runs are presented in Table \ref{table:runs}.

\begin{figure}
\centering
\includegraphics[height=7.5cm,angle=-90]{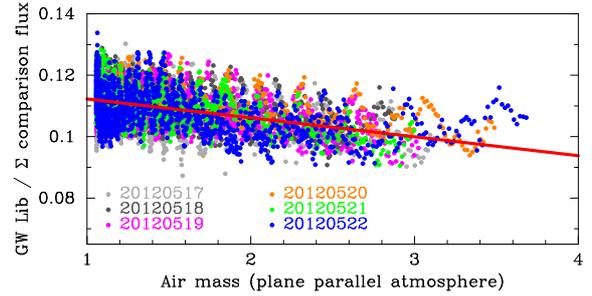}
\caption {The residual extinction in the photometry was modelled and removed
from the data by calculating the change in mean relative intensity with increasing
air mass. The times associated with each raw target / comparison flux ratio
were converted to an air mass (assuming a plane parallel atmosphere), and the
flux ratio was simultaneously fitted over the six nights for air masses 1 -- 3.
The mean intensity of GW Librae remains constant to a level much better than
the $\sim$10\% pulsation modulation, which appears in the plot as the large
scatter about the mean trend. These variations are averaged over $\sim$12 independent
measurements (two per night as the field rises and sets) to provide a much
more robust extinction correction than the standard per-night polynomial fit.}
\label{fig:airmassratio}
\end{figure}

\begin{figure*}
\centering
\includegraphics[width=0.95\textwidth]{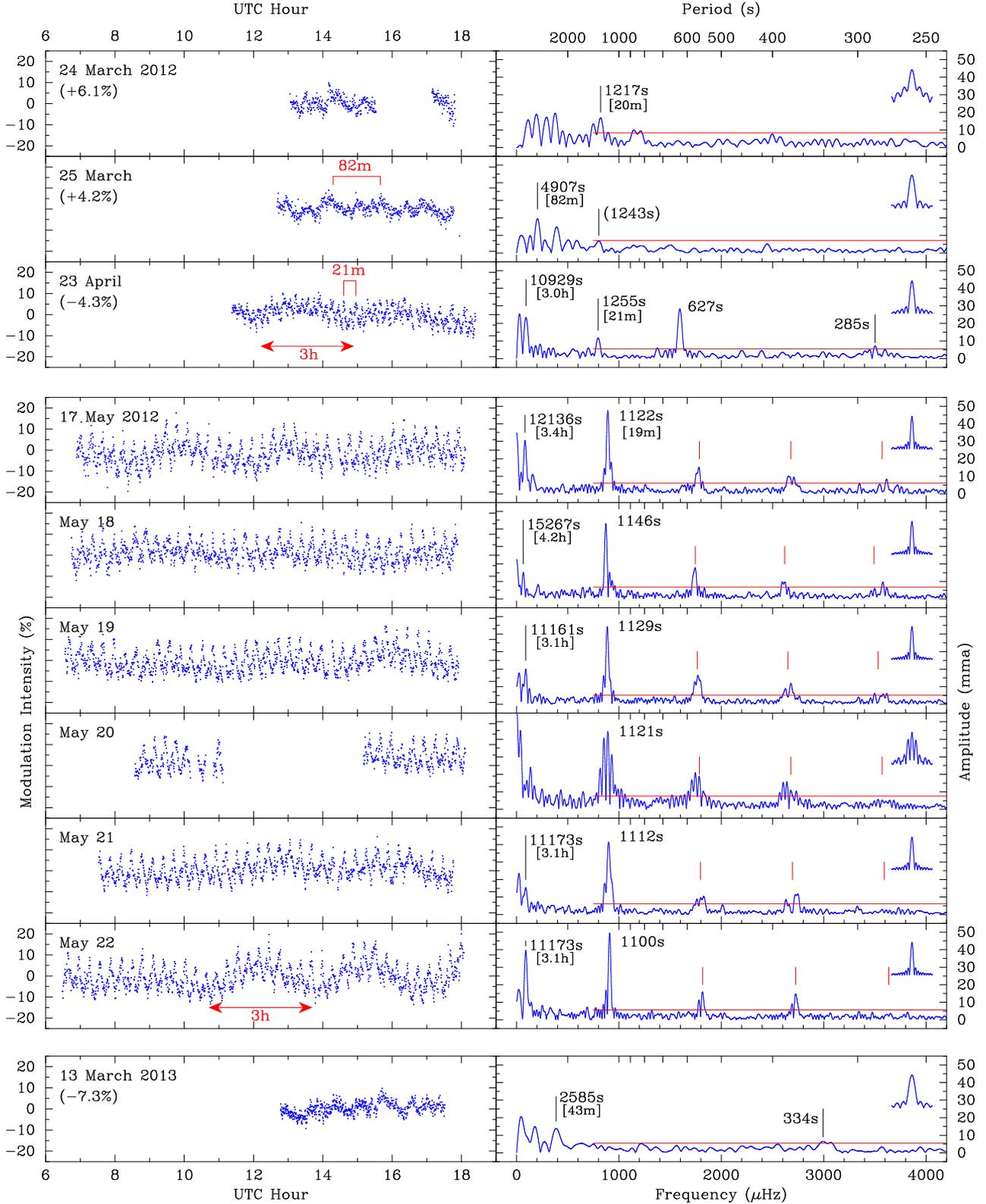}
\caption{
  The new observational data on GW Lib obtained in 2012 and March 2013 using
  the Puoko-nui photometer attached to the Mt John 1m telescope. All CCD
  exposure times were 30\,s. The panels on the left depict the flux
  variations about the mean levels in per cent modulation units, and the
  panels on the right show the corresponding DFTs for each data set in
  millimodulation amplitude (mma) units (10 mma = 1 per cent amplitude). The
  DFT window functions for each run are provided in the panel inserts
  and have the same horizontal frequency scale. The horizontal (red)
  lines in the DFT panels represent the False Alarm Probabilities
  (FAP) at the 1 in 1000 significance level (see text). The vertical
  (red) lines in the DFT panels indicate the harmonics at 2, 3, 4~$\times$
  the noted base frequency. The intensity of the six 2012 May runs are measured
  relative to a common mean value, which is $\sim$0.1 mag fainter (as measured through the BG40 filter)
  than the 2011 July observation campaign \citep{szkody12}. The percentages noted
  below the dates in the other four panels show the mean intensity difference of these
  runs relative to 2012 May.}
\label{fig:tsdft}
\end{figure*}

All observations were acquired using using the Puoko-nui high-speed photometer
\citep{chote14} with the 1m telescope at the University of Canterbury Mt John Observatory (UCMJO).
We used our standard BG40 broad B band filter in order to minimise the sky
contributions to the bluer WD flux. The photometer employs a frame transfer CCD,
so there were negligible deadtimes involved in the data acquisition. The CCD
frames were analysed and light curves extracted using our standard reduction
procedure \citep[see][]{chote14} with one change: the per-run polynomial fit
to correct for differential extinction is not appropriate for these runs because
GW Lib shows intrinsic variability on a similar timescale to the changing extinction ($\sim$hours), and it is
not possible to separate these effects within a single run. We instead take a
different approach, described in Fig.\ \ref{fig:airmassratio}, that fits an
extinction parameter simultaneously to the six 2012 May runs. The extinction
parameter from this fit was reused for the 2012 March, April and 2013 April runs.

Fig.\ \ref{fig:tsdft} presents the reduced time-series and amplitude Fourier
Transform (DFT hereon) plots of the new data. Similar plots of the 2011 runs are
presented in Figs. 12 and 14 of \cite{szkody12}. The horizontal (red) lines in the
DFT panels indicate the 0.1 per cent false alarm probabilities \citep[FAP;][]{sullivan08},
which are calculated using a Monte Carlo process and represent the threshold where
there is a 0.1 per cent probability that a peak in the DFT greater than this amplitude may
occur by a conspiracy of the random photometric noise. We calculate the FAP for each run
over the frequency range 750 -- 10000\,$\mu$Hz after prewhitening the 1100\,s, 280\,s, and 3\,h
signals discussed in the following sections. Due to the possible contributions
from 1/f noise and other effects our FAP evaluation procedures are less reliable
for the longer periods so limit the estimates to greater than 750\,$\mu$Hz.
The reality of the long period variations (e.g. the $\sim$3\,h modulations) are
best viewed in the time domain.

The two March 2012 runs show a $\sim$80 minute double-humped modulation, which
is longer than both the orbital period and the 78 minute superhump period
that was seen during the 2007 outburst \citep{kato08, bullock11}. The DFT
shows power near 20 minutes, but these peaks are not clearly distinguished.

The April 2012 light curve was dominated by a double-humped 1255\,s (21 minute)
periodicity superimposed on top of a longer three hour modulation which is
discussed in Section \ref{section:threehour}. The DFT also shows power in the
vicinity of the $\sim$280\,s pulsations seen during 2010 and 2011.

Less than one month later, on 2012 May 17, the 1255\,s signal appears
to have shifted to a shorter period of $\sim$1100\,s and roughly
doubled in amplitude. This was again superimposed on top of a 3\,h
modulation. The DFT of this first run showed a clear signal at 1120\,s and
its harmonics, but the peak profile appeared to indicate some underlying
frequency structure.

The coordinates of GW Lib (15$^{\rm{h}}$19$^{\rm{m}}$55$^{\rm{s}}$,
$-25^\circ00^\prime25^{\prime\prime}$) allow it to be observed all
night from Mt John in May. This coincided with a run of exceptionally
good observing conditions, and enabled a total of 61\,h of excellent
quality photometry to be measured over 6 nights -- an impressive 42 per cent
observing duty cycle for these single-site runs.

Our next observation of GW Lib was not until March 2013, and by this time GW
Lib had returned to a similar state as was seen in March 2012. A short run
on 2012 June 24 using the Agile photometer on the 3.5\,m telescope at Apache
Point Observatory confirmed that the $\sim$1100\,s modulation remained strong
(P. Szkody, private communication). We are not aware of any other interim
observations, so our best constraint on the length of time that the
modulation was visible is three months to one year.

The 21 minute modulation was detected again during an observing campaign
in April -- May 2015. The analysis of these new data is ongoing, and results
will be detailed in a future publication.

\section{The 1100\,s Periodicity}

The dominant feature of the May 2012 runs was a double-humped $\sim$1100\,s
(19 minutes) modulation. The DFTs for each night (Fig.\ \ref{fig:tsdft}; right
panels) show an apparent frequency variation of tens of seconds between nights,
and the April 23 run suggests that the variability range may be even larger ($\sim$100\,s).

This appears to be the return of a modulation that was visible for at least four
months in 2008 \citep{copperwheat09, schwieterman10, bullock11, vican11}. The
shape and amplitudes have changed from the earlier observations: in 2012 we find
a higher-amplitude ($50\,$mma versus $22\,$mma) and non-sinsoidal double-humped
modulation profile.

The origin of this period was not clear from the 2008 observations:
\cite{copperwheat09} suggested that it could arise from accretion-based
phenomena originating in the disk, but \cite{vican11} countered this by
explaining that such quasi-periodic oscillations \citep{warner03} are a
characteristic of high accretion rate systems and aren't expected to occur in
GW Lib. Instead they proposed that it may be a newly driven pulsation mode of
the WD. The pulsation mode interpretation is strongly supported by our observations
that show the modulation return in 2012 with a very similar period and behaviour.

\begin{figure}
\centering
\includegraphics[width=7.5cm,angle=-90]{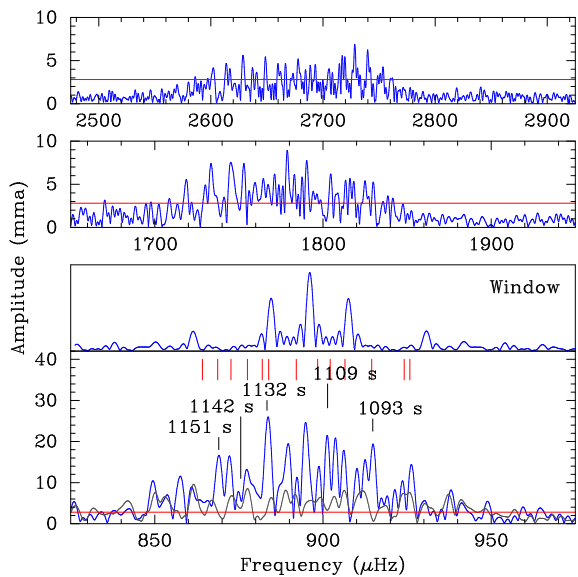}
\caption {
  The combined DFT of the May 2012 runs is not consistent with stable
  pulsation modes. The DFT power in the vicinity of the main 19 minute
  periodicity (bottom panel) is dominated by five periods on top of a hump
  of less coherent power that requires a thirteen frequency fit (vertical red lines)
  to reduce below the 0.1 per cent FAP threshold (horizontal red line). The window
  function (upper-bottom panel; see text) shows the signature of the sampling
  aliases introduced by the daily gaps in the light curves. The harmonics
  (top panels) do not show any clear frequency structure.}
\label{fig:dft_may12}
\end{figure}

It is a standard practice when observing the stable WD pulsators to combine
multiple nights of photometry to improve the frequency resolution of the DFT.
We attempt this for our May 17 -- 22 observations in Fig.\ \ref{fig:dft_may12}. 

The main panel at the bottom of this figure presents the amplitude
spectrum (blue) around the main periodicity, which resolves a messy comb of peaks
that can be roughly fitted as the sum of five (simultaneously fitted) stable
sinusoidal modulations. There is not a one-to-one
correspondence between frequency peaks and periods because the daily gaps between
runs creates an artificial modulation that introduces aliases separated by
11.6\,$\mu$Hz (1 day$^{-1}$). The signature of these aliases is illustrated
by calculating the DFT of a noise-free sinusoid that is sampled at the the same
times as the original data. This DFT \emph{window} function is plotted for a
1100\,s sinusoid immediately above the bottom panel in Fig.\ \ref{fig:dft_may12}. 

The five-frequency fit does a poor job of reproducing the phases of
the individual nightly light curves, and the amplitude shows beating effects that
are not consistent with the data. These appear as a hump of power extending above
the 0.1 per cent FAP threshold in the DFT of the residual signal (bottom panel; grey).
A simultaneous thirteen-frequency fit (indicated with red lines in Fig.\ \ref{fig:dft_may12})
is required to adequately describe the light curves in the time domain, but large
beating effects remain visible in the gaps between runs. This is clearly not a realistic model.

These results show that the frequency variability of the modulation cannot be
attributed to beating between a set of stable pulsation modes, so we conclude that
the period(s) is/are truly variable in frequency and/or amplitude. This presents a
problem for the standard model of pulsations that is applied to the slowly rotating
isolated WDs -- frequencies should change only over evolutionary timescales, not over hours.
We know, however, that this assumption is not strictly true even for the isolated WDs:
the cool DAV stars such as G29-38 \citep{kleinman98} and the DBV star GD\,358 \citep{kepler03}
show amplitude modulation, with power that shifts between different stable modes across
observing seasons. The recent discovery of flaring DAV WDs \citep{bell15, hermes15}
provide a new class of WD pulsator that shows power shifting between modes in response
to flare events that regularly repeat on a timescale of a few days.

\subsection{Spectrogram analysis}\label{section:subnight}

A useful technique for visually depicting a multi-periodic signal that
changes over time is the spectrogram (also known as a running Fourier
transform). Spectrograms show a running plot of the DFT amplitude
using colour over a range of frequencies against time. They are
commonly used in audio signal analysis, but have also been applied to the
study of pulsating WDs that exhibit amplitude variability
\citep[e.g.][]{hermes15, provencal12}.

The spectrogram extends the idea of dividing a signal into discrete
chunks for comparison into a continuous process: each vertical slice
in a spectrogram plot corresponds to the DFT amplitude calculated using
data within a specified time window. The window is moved continuously
across the data set in order to generate the plot. There is a trade
off between frequency and time resolution: we chose a spectrogram
window size of 3\,h to give a reasonable time
resolution while not sacrificing too much frequency resolution.

\begin{figure*}
\centering
\includegraphics[width=\textwidth,angle=0]{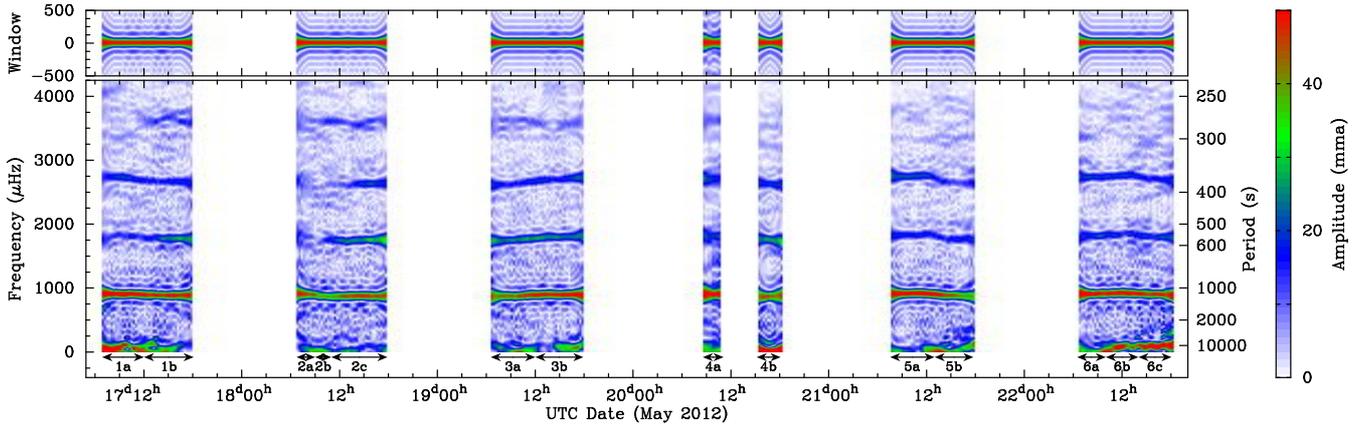}
\caption{
  A spectrogram of the May 2012 runs shows how the modulation periods evolved
  over six consecutive nights in May 2012. The DFT amplitude of a three-hour
  segment of the data is plotted (using the colour-coded scheme given at the
  right of the figure) as a function of frequency and mid-segment time.
  The frequency resolution is indicated by the top panel which shows the
  spectrogram of a noise-free sinusoid sampled at the same times as the data
  (a direct analogue to the DFT window function).}
\label{fig:spectrogram}
\end{figure*}

A spectrogram of the May 2012 data is presented in Fig.~\ref{fig:spectrogram}.
No significant signals were found at frequencies above 4200$\,\mu$Hz (periods
shorter than 240\,s), and so this was chosen as the maximum frequency limit of the plot.

The harmonic frequencies are immediately apparent in the spectrogram at multiples
of the strong fundamental frequency. The fundamental frequency can be seen to change
over the course of each night, and this change is even more visible in the harmonics
where the change is multiplied by the harmonic order. The first three runs show
additional power near 285\,s (3500$\,\mu$Hz) that does not follow the pattern of the
1100\,s modulation; this appears to be the 280\,s pulsation mode that was seen through
the 2011 observations.

The runs on May 17, 18, and 21 appear to show step-changes in frequency
(the step itself is blurred over the 3h window), while the May 19 and 22 runs
show a more continuous wandering.

In an effort to obtain more quantitative results,
the spectrogram was used as a guide to split each nightly run into two or three
sub-runs, which were then individually reanalysed. Table \ref{table:periods}
presents the details and periods found in each of these sub-runs. 

\begin{table*}
\label{table:periods}
\centering
\caption{A summary of the complete May 2012 data set for GW Lib and
  the various periods deduced from nightly DFTs and multi-frequency
  fits to various sections of the time-series data.}
\begin{tabular}{ccccccccc}
\hline
UT Date & Sub & Time span & $\Delta$t & Period 1  & Ampl. 1    & Period 2        & Ampl. 2    \\
(2012)  & run &   (UTC)   &   (h)     &   (s)     & (mma)      & (s)             & (mma)      \\
\hline
17 May & 1a & 06:53 -- 12:00 & 5.1 & 1108 $\pm$ 1 & 52 $\pm$ 2 &                 &            \\
       & 1b & 12:00 -- 18:06 & 6.1 & 1133 $\pm$ 1 & 50 $\pm$ 2 & 276.8 $\pm$ 0.3 & 25 $\pm$ 2 \\
18 May & 2a & 06:43 -- 09:00 & 2.3 & 1114 $\pm$ 18& 52 $\pm$ 3 &                 &            \\
       & 2b & 09:00 -- 11:00 & 2.0 & 1176 $\pm$ 80& 33 $\pm$ 2 & 280 $\pm$ 1     & 14 $\pm$ 3 \\
       & 2c & 11:00 -- 17:52 & 6.9 & 1150 $\pm$ 1 & 46 $\pm$ 1 & 280.5 $\pm$ 0.2 & 11 $\pm$ 1 \\
19 May & 3a & 06:33 -- 12:00 & 5.5 & 1145 $\pm$ 1 & 46 $\pm$ 1 & 277.0 $\pm$ 0.3 & 16 $\pm$ 1 \\
       & 3b & 12:00 -- 17:55 & 5.9 & 1119 $\pm$ 1 & 50 $\pm$ 1 &                 &            \\
20 May & 4a & 08:33 -- 11:07 & 2.6 & 1110 $\pm$ 1 & 57 $\pm$ 2 &                 &            \\
       & 4b & 15:10 -- 18:06 & 2.9 & 1146 $\pm$ 1 & 44 $\pm$ 2 &                 &            \\
21 May & 5a & 07:32 -- 13:00 & 5.5 & 1096 $\pm$ 1 & 51 $\pm$ 1 &                 &            \\
       & 5b & 13:00 -- 17:46 & 4.8 & 1137 $\pm$ 1 & 47 $\pm$ 2 &                 &            \\
22 May & 6a & 06:28 -- 10:00 & 3.5 & 1105 $\pm$ 1 & 51 $\pm$ 2 &                 &            \\
       & 6b & 10:00 -- 14:00 & 4.0 & 1092 $\pm$ 1 & 53 $\pm$ 1 &                 &            \\
       & 6c & 14:00 -- 18:03 & 4.0 & 1113 $\pm$ 1 & 48 $\pm$ 2 &                 &            \\
\hline
\end{tabular}
\end{table*}

The period uncertainties were calculated using a Monte Carlo procedure: first the
time-series data were pre-whitened by removing all the significant frequencies,
a noise histogram characterising the residual variations was created, and this
was fitted with a Gaussian. A synthetic lightcurve was computed using the modelled
periods and random noise sampled from the fitted Gaussian, and then a new period
determination for a specific period was made using a non-linear squares fit while
holding the other periods fixed.  Repeating this procedure 1000 times yielded
histograms for the fitted period and amplitude values.  Fitted Gaussian parameters
from these histograms were used to obtain the stated period and amplitude uncertainties.

We note that this procedure measures only the uncertainty in fitting the data,
and does not account for any systematic uncertainties that may be present.
As such, they should be treated as lower limits.

\begin{figure*}
\centering
\includegraphics[height=\textwidth,angle=-90]{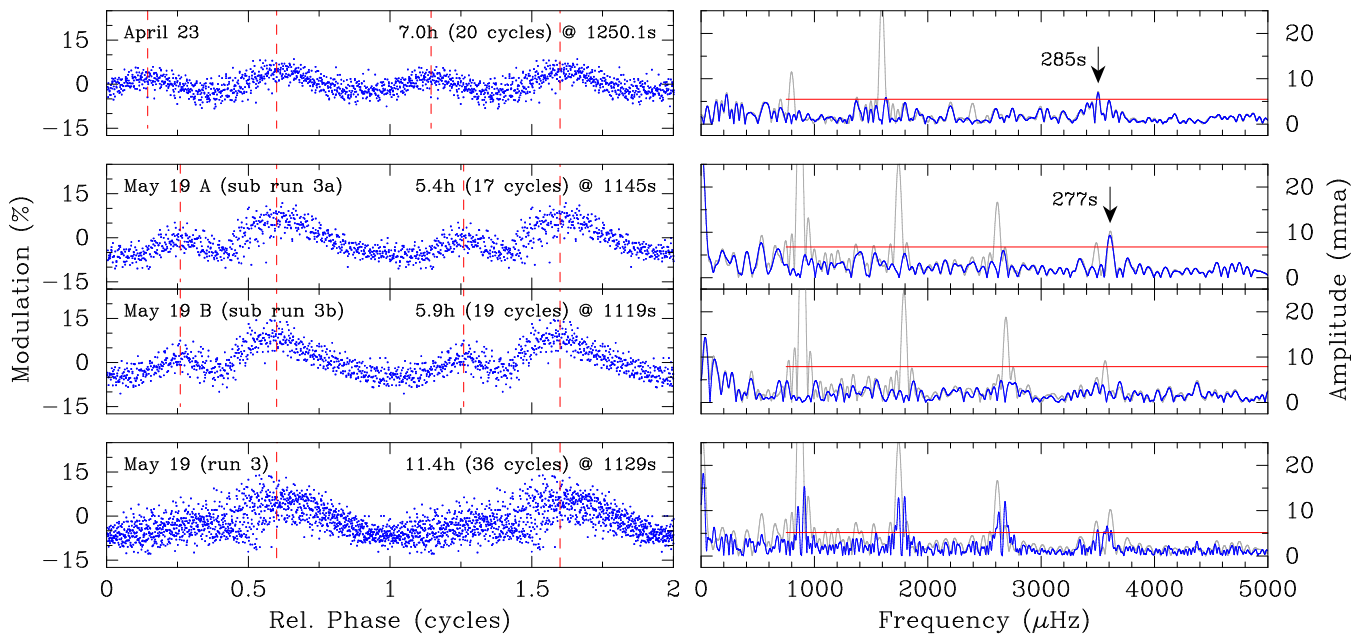}
\caption {The short-term coherency of the 1100\,s modulation is explored by
  presenting folded light curves (left panels) and DFTs (right panels) of the
  2012 May 19 and April 23 runs. The $\sim$3\,h modulation was first prewhitened
  from each run, and the data were then folded at the best-fit $\sim$1100\,s period.
  The vertical lines in the folded light curve panels indicate the (arbitrarily
  aligned) phases of the modulation peaks. The right panels present DFTs before
  (grey) and after (blue) pre-whitening the best-fit modulation period indicated
  in the left panels. The horizontal lines in the DFT panels indicate the
  standard 0.1 per cent FAP threshold. See text for an interpretation of these plots.}
\label{fig:folddft}
\end{figure*}

We find that the sub-runs are each well described by a mono-periodic non-sinusoidal
modulation, and that four of the sub-runs feature the 280\,s pulsation mode.
Fig. \ref{fig:folddft} demonstrates the improvements that come from
considering the individual sub runs by presenting light curves folded on the
$\sim$1100\,s periods (left panels), and DFTs (right panels) for the two May
19 sub-runs (bottom panels) and the full May 19 and April 23 runs (top panels)
for comparison.

In both the April and May runs we see a clear double-humped profile, but the
shape and relative phase of the humps differ. The profile shape does not
show significant changes between the May sub-runs, but the best-fit periods
change by as much as 30 seconds.

The bottom panel shows the result of folding the full May 19 light curve on its
best-fit period. The phase profile is poorly defined compared to the two sub-runs,
and there are significant residuals in the prewhitened DFT. This is quite a
contrast to the sub-runs that it covers, which are individually more coherent
and appear in their DFTs as a single well defined non-sinusoidal period.

\section{The 280 Second Periodicity}\label{section:shortmode}
Figure \ref{fig:folddft} shows an additional result: the $\sim$280\,s pulsation
is well defined in the May 19 1a sub-run, but indistinguishable from the noise in
the 1b sub-run.  Similar sharp changes are seen in the May 17 and 18 runs.

\begin{figure*}
\centering
\includegraphics[width=\textwidth]{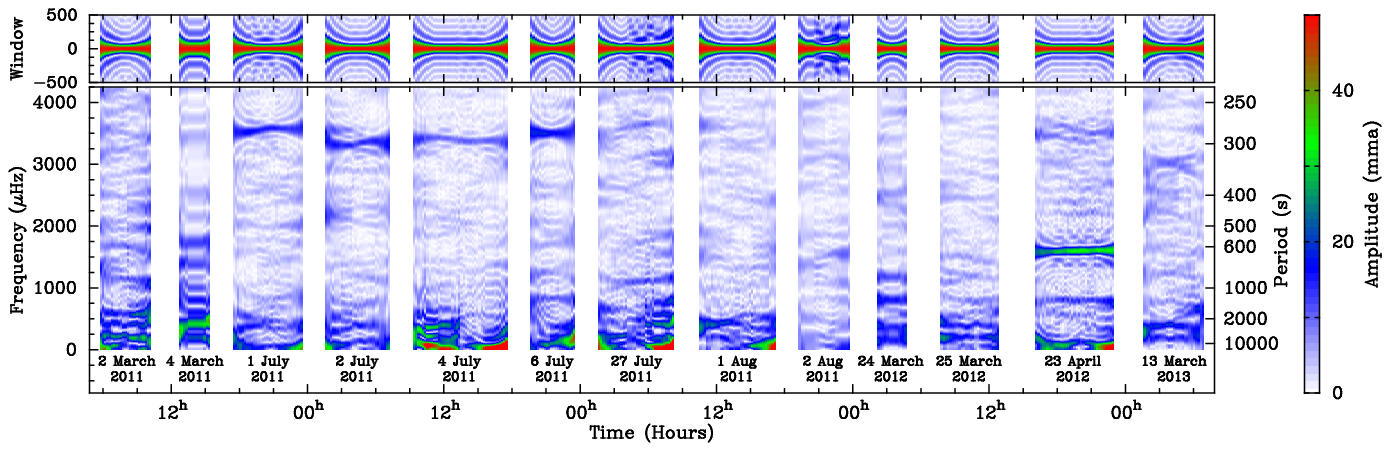}
\caption {Spectrograms of the other Mt John runs reveal
a similar wandering behaviour within the $\sim$280\,s pulsation
(near $3400\,\mu$Hz). The disjoint runs have been combined on to a
common time baseline for display purposes.}
\label{fig:2011spectrogram}
\end{figure*}

This suggests that the frequency multiplets reported in \cite{szkody12} are
likely to be artefacts caused by amplitude and/or frequency variability.  We
test this idea by running a spectrogram analysis over our 2011 Mt John data,
which is presented in Fig \ref{fig:2011spectrogram}.

The 2011 July 1, 2, 4 runs show similar wandering behaviour to the May 2012 1100\,s
periodicity: power appears to shift between temporarily
stable frequencies. The 280\,s mode is visible in the 2011 July 6 run, but the run
length is too short to detect any changes.  We see power in the expected region
for the April 2012 run, but it appears to have relatively poor coherency.

\section{The 3 Hour Periodicity}\label{section:threehour}
A $\sim$3\,h modulation is visible by eye in the April and May 2012 light
curves (Fig. \ref{fig:tsdft}), but the specific details of the modulation are
masked by the $\sim$19\,minute signal.

\begin{figure}
\centering
\includegraphics[height=8cm,angle=-90]{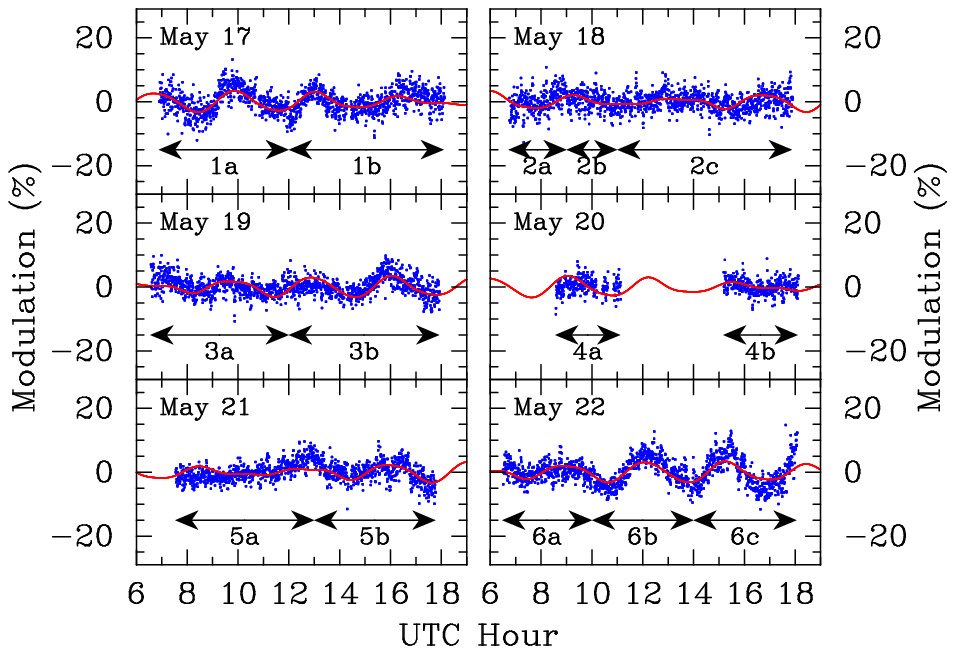}
\caption{Pre-whitening the $\sim$19\,minute and $\sim$280\,s signals from the
May 2012 light curves provides a clearer look at the $\sim$3\,hour modulation.
The best two-frequency model is shown in red, and the sub-run divisions are indicated
below the data.}
\label{fig:longperiod}
\end{figure}

Figure \ref{fig:longperiod} presents pre-whitened light curves that provide a
clearer view of the long period modulation. We find that this variability can be
broadly described by beating between a 3.4\,hour and a 2.8\,hour signal, but,
like the $\sim$19\,minute modulation, significant residuals remain after
prewhitening this model fit.  We therefore consider it more likely that this
behaviour is caused by a single quasi-stable modulation.

We find that our sub-run divisions (which were chosen empirically based
on the behaviour of the $\sim$19\,minute modulation) appear to coincide with
discontinuities in this long period signal: the 1a/1b and 6b/6c run splits
occur immediately before sharp increases in the modulation, and the 5a/5b and 6a/6b
splits line up immediately before sharp decreases in the signal.

\begin{figure}
\centering
\includegraphics[height=8cm,angle=-90]{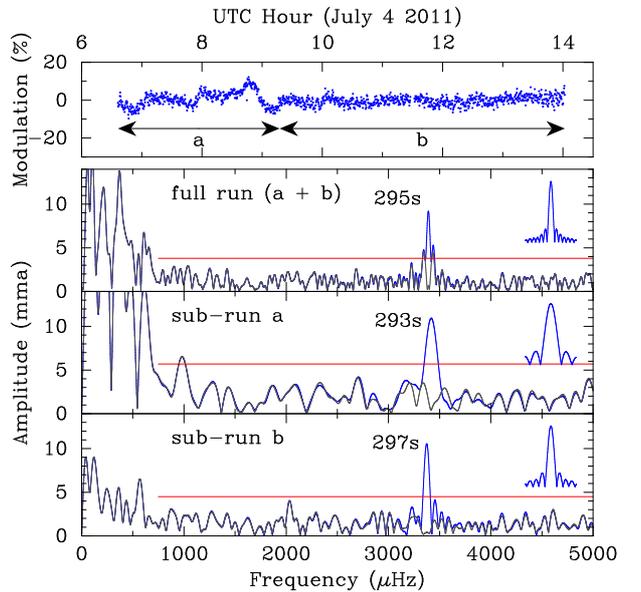}
\caption{The 2011 July 4 run is split into two sub-runs at the point where
the long-period variability changes character.  The blue curves in the bottom
three panels show the DFTs of the three data sets, and the grey curves show the DFTs
after prewhitening the noted periods.  The horizontal red lines show the 0.1\%
FAP thresholds.  We find different best-fit periods for the $\sim$280\,s pulsation
mode in each sub run, and note that these periods cleanly prewhiten from the data.
The DFT of the full run appears to show side-peaks around the best-fit period,
which is a sign of the variability of the mode.}
\label{fig:july2011dft}
\end{figure}

The spectrogram of the 2011 July 4 run (Fig. \ref{fig:2011spectrogram}) shows a
sharp change in the long-period signal that appears to be correlated with the
change in the 280\,s modulation.  The DFT of the full run shows an apparent
frequency triplet centred on 295\,s, but when we split the night into two
sub-runs based on the behaviour of the long-period modulation we find different
best-fit periods of 293\,s where the long period modulation is strong, and 297\,s
where it is weak (see Fig \ref{fig:july2011dft}).

It is possible that these correlations are coincidental, but we consider it
unlikely. Another example of this correlation can be seen in Fig. 2 of \cite{toloza15},
which shows a UV light curve of GW Lib obtained over three HST orbits in 2013.
Their light curve of the second orbit shows the 280\,s pulsation appear with
increasing amplitude on the rising slope of a strong long-period modulation.
\cite{schwieterman10} also reported an apparent anti-correlation between the
amplitude of the 2.1 hour and 19\,minute signals in 2008.  These results
together suggest that a common mechanism is acting to affect the $\sim$280\,s,
$\sim$19\,minute, and long period variability in GW Lib.

\section{Conclusions}

Observations of GW Librae over a three month period in 2012 showed the return of
a $\sim$1100\,s modulation period that was previously detected over several months
in 2008. The fact that this returned with an essentially identical period, and
its extended near-coherent modulation behaviour rule out the earlier speculation that
this was a transient or quasi-periodic oscillation associated with the accretion disk.

We find that the frequency variation cannot be described in terms of beating between
stable pulsation modes, and that the period of a single mode appears to switch between
quasi-stable values every few hours. Observations in March and April 2012 constrained
the overall growth time of this modulation to less than two months, and suggested that
the modulation showed large changes in period and the relative phase of the two humps
during its growth.

We also track the behaviour of the $\sim$280\,s pulsation mode where it is visible,
and the long-period $\sim$3\,hour modulation. We find that changes in the
behaviour of these modulations appear to be correlated in time with changes in
the behaviour of the $\sim$1100\,s signal. We show a similar correlation in
our previously published 2011 July 4 run, and note similar correlations that have
been presented in two other papers.

The 280\,s mode has been clearly identified as pulsation behaviour on the WD
\citep{szkody12}, and we can therefore infer that the $\sim$1100\,s periodicity
must also be a pulsation mode. This allows us to revise our estimate of when
pulsations returned after the 2007 outburst from 2010 back to 2008.

The $\sim$3\,hour modulation period has not been previously reported, but the
modulation amplitude and quasi-periodic behaviour is consistent with the known
2/4 period modulation.  We therefore believe that this is another state of the
same modulation phenomenon for which we still lack a physical explanation.

The origin of the sharp changes in the pulsation behaviour remains a mystery,
but there is observational evidence to suggest that this phenomenon is a general
feature of the accreting pulsators:

\begin{itemize}
	\item SDSS 0745+4538
\citep{mukadam07} is another pulsating CV system, which
features a double-humped $\sim$1100\,s period that was found to
change by as much as 30\,s between nights. This pulsation
disappeared when the system went through an outburst in 2006, and
later returned with the same periods and behaviour in 2010 \citep{mukadam11}.
\item An extensive photometric data set on V455 And \citep{szkody13b} following
its 2007 outburst reveals pulsation periods that drift over a wide range
of values within certain allowed regions (B. G{\"a}nsicke, private
communication).
\item The three canonical pulsation periods that GW Lib
showed before its 2007 outburst were also found to be unstable at the
$\sim\mu$Hz level \citep{vanzyl04}.
\end{itemize}

These period transitions pose a challenge for observing these systems, as it
means that we cannot improve the signal to noise ratio in the DFT by extending
the observation baseline -- this simply introduces artefacts caused by the
frequency and amplitude modulation. The spectrogram technique presented here
provides a useful method for monitoring the stability of the
pulsations if sufficiently high quality photometry is available.

Recent models of rapidly rotating pulsating WDs \citep{townsley16} suggest that
retrograde pulsation modes with periods close to the WD spin period may be slowed
down to an apparent period of $\sim$hours in the observer's frame of reference. This
provides a convenient explanation for both the temporal correlation in the
behavioural changes, and also for how the long-period mode can change between
the observed 2/3/4 hour periods (a small wandering in the true period would be
magnified to a large change in the visible period).

An alternative explanation for the correlation is that the long period
signal is tracking large-scale changes in the WD atmosphere, which are then also
perturbing the frequency and amplitudes of the observed pulsation modes.

The theory that the long period modulation is associated with the accretion
disk appears to be ruled out by \cite{toloza15}, who show that the modulation seen
during the 2013 HST run was associated with temperature changes in the WD atmosphere.

The recent discovery of isolated DAV pulsators that undergo regular
flare events \citep{bell15, hermes15} may provide some insights into this behaviour.
The pulsation modes that are visible during the flares have much higher
amplitude and shorter period than the quiescent modes, but then all
pulsation modes rapidly cease for a time, before growing back with the
original quiescent periods. One proposed mechanism for this behaviour
is that nonlinear coupling between pulsations cause power to be dumped
into damped daughter modes. Similar nonlinear effects could be
affecting the behaviour that we are seeing in GW Lib.

While we are making progress on untangling the observed behaviour of GW Lib and
the other accreting pulsating WD systems, many questions and uncertainties remain.
Further monitoring and modelling is required to sort out the details.

\section*{Acknowledgments}
We thank the Marsden Fund of NZ for providing financial support for
this research and the University of Canterbury for the allocation of
telescope time for the project. PC acknowledges funding from the
European Research Council under the European Union's Seventh Framework
Programme (FP/2007-2013) / ERC Grant Agreement n. 320964 (WDTracer).
We also thank Boris G{\"a}nsicke and the anonymous referee for
insightful feedback and discussions that improved the manuscript.

\bsp
\bibliographystyle{mnras}

\label{lastpage}

\end{document}